\documentclass{aastex}

\shortauthors{Lee et al.}
\shorttitle{Improved Estimation of LBG Physical Parameters from Photometry}

\begin{document}

\title{Improving the Estimation of Star formation Rates and Stellar
Population Ages of High-redshift Galaxies from Broadband Photometry}

\author{Seong-Kook Lee}
\affil{Department of Physics and Astronomy, Johns Hopkins University,
3400 North Charles Street, Baltimore, MD 21218-2686, USA}
\email{joshua@pha.jhu.edu}

%
\author{Henry C. Ferguson}
\affil{Space Telescope Science Institute, 3700 San Martin Drive,
Baltimore, MD 21218, USA}

\author{Rachel S. Somerville}
\affil{Space Telescope Science Institute, 3700 San Martin Drive,
Baltimore, MD 21218, USA}

\author{Tommy Wiklind}
\affil{Space Telescope Science Institute, 3700 San Martin Drive,
Baltimore, MD 21218, USA}

\and

\author{Mauro Giavalisco}
\affil{Astronomy Department, University of Massachusetts,
Amherst, MA 01003, USA}

\begin{abstract}

We explore methods to improve the estimates of star formation rates and
mean stellar population ages from broadband photometry of high-redshift
star-forming galaxies.
We use synthetic spectral templates with a variety of simple parametric
star formation histories to fit broadband spectral energy distributions.
These parametric models are used to infer ages, star formation rates, and
stellar masses for a mock data set drawn from a hierarchical semi-analytic
model of galaxy evolution.
Traditional parametric models generally assume an exponentially declining
rate of star formation after an initial instantaneous rise.
Our results show that star formation histories with a much more gradual
rise in the star formation rate are likely to be better templates, and are
likely to give better overall estimates of the age distribution and
star formation rate distribution of Lyman break galaxies (LBGs).
For $B$- and $V$-dropouts, we find the best simple parametric model to be one
where the star formation rate increases linearly with time.
The exponentially declining model overpredicts the age by $100 ~\%$ and $120 ~\%$
for $B$- and $V$-dropouts, on average, while for a linearly-increasing model, the
age is overpredicted by $9 ~\%$ and $16 ~\%$, respectively.
Similarly, the exponential model underpredicts star formation rates by
$56 ~\%$ and $60 ~\%$, while the linearly increasing model underpredicts by $15 ~\%$ and
$22 ~\%$, respectively.
For $U$-dropouts, the models where the star formation rate has a peak (near $z \sim 3$)
provide the best match for age --- overprediction
is reduced from $110 ~\%$ to $26 ~\%$ --- and star formation rate --- underprediction
is reduced from $58 ~\%$ to $22 ~\%$.
We classify different types of star formation histories in the semi-analytic
models and show how the biases behave for the different classes.
We also provide two-band calibration formulae for stellar mass and
star formation rate estimations.

\end{abstract}


\section{Introduction}


A widely used method to estimate various physical parameters of
high-redshift galaxies --- such as stellar mass, star formation rate
(SFR), stellar population mean age, and dust extinction --- is
to compare their observed photometric spectral energy distributions
(SEDs) with spectral templates from stellar population synthesis
models, such as \citet[][hereafter BC03]{bru03} models \citep[e.g.,][]{saw98,pap01,sha01}

There have been several investigations of the uncertainties in
these ``SED-fitting methods" arising from various sources,
including the assumed form of the initial mass function
\citep[IMF; e.g.,][]{pap01,con09}, extinction law
\citep[e.g.,][]{pap01}, and the treatment of stellar evolution
\citep[e.g.,][]{mar06,con09}.

Our focus in this paper is on star-forming galaxies at redshifts $z > 3$.
Recently, \citet[][hereafter L09]{lee09} have shown that there can be
significant systematic biases in estimates of ages and SFRs of
high-redshift Lyman break galaxies (LBGs) through the SED-fitting methods,
even when the above-mentioned uncertainties are minimal.
L09 have also investigated the causes of these biases, one of which is
the difference between the galaxies' actual star formation histories
(SFHs) and those assumed in the SED-fitting procedure --- in other
words, oversimplified assumptions about the galaxies' SFHs in the
SED-fitting.
Exponentially declining SFRs are typically assumed
in SED-fitting, whereas the SFHs of real galaxies are probably
more complex.

Motivated by our previous work in L09, and considering the importance of
robust estimation of various physical parameters in the study of galaxy
formation and evolution, we continue our analysis of the uncertainties and
biases in SED-fitting by considering models wherein the peak in the
SFR is not at the beginning.
The standard exponentially decaying SFRs are a relic of monolithic models
of galaxy formation, and have very little relevance to hierarchical models.
They have been used in SED-fitting for convenience, but it seems worth
investigating whether some other model with a small number of free parameters
could do a better job of providing estimates of ages, SFRs, and stellar masses.
The goal is to make use of all the photometric information (rather than
just a UV slope and IR luminosity), while at the same time making very
few physical assumptions about galaxy evolution.

This study, like L09, is based on the analysis of SEDs and SFHs of
model LBGs from the semi-analytic models of galaxy formation.
We compare the intrinsic values of physical parameters of model LBGs extracted
from the models to the values derived through the SED-fitting methods.

The models are explained in Section 2, and results of SED-fitting are
presented in Section 3. In Section 4, we present the derivation of stellar mass,
SFR, and age from luminosity and color, and in Section 5, we summarize the results.
Throughout the paper, we adopt a flat $\rm{\Lambda}$CDM cosmology,
with ($\Omega_{m}, \Omega_{\Lambda}$) = (0.3,0.7) and $H_{0}$ =
70 \rm{$km$ $s^{-1}$ $Mpc^{-1}$}.

\section{Lyman Break Galaxy Samples from the Semi-analytic Models}

Semi-analytic models of galaxy formation are based on the theory of the
growth and collapse of fluctuations in a $\Lambda$CDM initial power spectrum
via gravitational instability.
The models used here are based on the dark matter halo merger tree of
\citet{som99} and include simplified analytic treatments for
various physical processes, including star formation, galaxy mergers, chemical
evolution, and feedback effects from various sources.
For star formation recipes, a quiescent mode (governed by a Kennicutt-like
law) as well as a burst mode (driven by galaxy mergers) are included.
We assume that a constant fraction of metals are produced per unit stellar
mass.
Thus, our semi-analytic models incorporate realistic
star formation and chemical enrichment histories that are
motivated by the hierarchical scheme of structure formation. Dust
extinction is modeled with the assumption that the face-on
optical depth in the $V$-band is given as $\tau_{V,0} =
\tau_{dust,0} \times (\dot{m_{*}})^{\beta_{dust}}$, where
$\tau_{dust,0}$ and $\beta_{dust}$ are free parameters, which are set
as 1.2 and 0.3, to match the observations in the GOODS-S.
The Calzetti attenuation curve \citep{cal00} is used in calculating dependence
of the extinction on wavelength.

In this work, we use the same model run which is used in L09.
This version of the model is similar to the models used in \citet{som01} and
\citet{idz04}, and has been shown by these authors to reproduce
many observed properties reasonably well --- such as number densities, luminosity functions, and
color distributions --- of high-$z$ LBGs in the GOODS-S field as well as the
evolution of global SFR density and of metal enrichment.

Model $U$-, $B$-, and $V$-dropout galaxy samples are selected using the same LBG
color-selection criteria as in L09 (Equations (1)-(9)).
Samples are further trimmed by applying the ACS-$z$ band magnitude limit
($z_{850} \leq 26.6$) of the GOODS-S field as in L09.
The mean redshifts of these model dropout samples are $z \sim$ 3.4, 4.0, and
5.0, respectively.

\section{Spectral Energy Distribution Fitting} \label{sedfit}

\subsection{Assumed SFHs and the Results of SED-fitting} \label{fitresult}

Motivated by the systematic biases in SED-fitting with widely used,
exponentially declining SFHs assumed, here we
try SED-fitting analysis with alternative SFHs
--- ``delayed" SFHs --- incorporated.

The SFR, $\Psi (t,\tau)$, in the delayed SFHs is given as

\begin{equation}
\Psi (t,\tau) \varpropto \frac{t}{\tau^{2}} e^{-t / \tau},
\end{equation} \label{delaysfr}

where $t$ is the time since the onset of star formation and $\tau$ is the
star formation time scale parameter.
At small $t$, the SFR increases, dominated by the linear
term, and peaks at $t = \tau$.
It starts to decrease after $t= \tau$, and is eventually dominated by the
exponential term at large $t$.
Thus, with very small $\tau$, the delayed SFHs are similar to the
generally used, exponentially declining SFHs, and with very large $\tau$,
they are nearly linearly increasing SFHs.

We vary the values of $t$ and $\tau$ within different allowed ranges
for SED-fitting with this form of delayed SFH, including :
(1) varying $\tau$ from 0.2 to 2.0 Gyr,
(2) limiting $\tau$ to be $\sim 0.9 \times t$, and
(3) setting  $\tau = 10.0$ Gyr.
Version (3) results in linearly increasing SFHs for
the high-redshift galaxies analyzed here ($3 \lesssim z \lesssim 5.6$),
because it peaks after 10 Gyr since the onset of star formation.

In semi-analytic models as well as in SED-fitting, we adopt BC03
models with a \citet{cha03} IMF.
The \citet{cal00} attenuation curve and the \citet{mad95} law are used for
dust extinction inside a galaxy and for intergalactic extinction due to
neutral hydrogen, respectively.
Internal dust extinction is parameterized through the color excess, $E(B-V)$,
with values from 0.0 to 0.95 in a step size of 0.025.
Similarly as in L09, metallicity is treated as a free parameter with two sub-solar
(0.2 and 0.4 $Z_{\sun}$) and solar metallicities allowed.

For SED-fitting, fluxes in 11 broad bands --- ACS $B_{435}$, $V_{606}$, $i_{775}$,
$z_{850}$ bands, ISAAC $J,~H,~Ks$ bands, and IRAC 3.6, 4.5, 5.8, 8.0 $\mu$m
channels --- are used (same as in L09).
And, from the observed LBGs in GOODS-S, mean errors for different magnitude
bins at each passband of ACS, ISAAC, and Infrared Array Camera (IRAC) are calculated, and assigned to
each SAM galaxy photometric value according to their magnitudes in the
calculation of $\chi^{2}$ --- thus our model LBGs have similar signal-to-noise
ratio (S/N) with observed GOODS-S LBGs.

\begin{figure}
\plotone{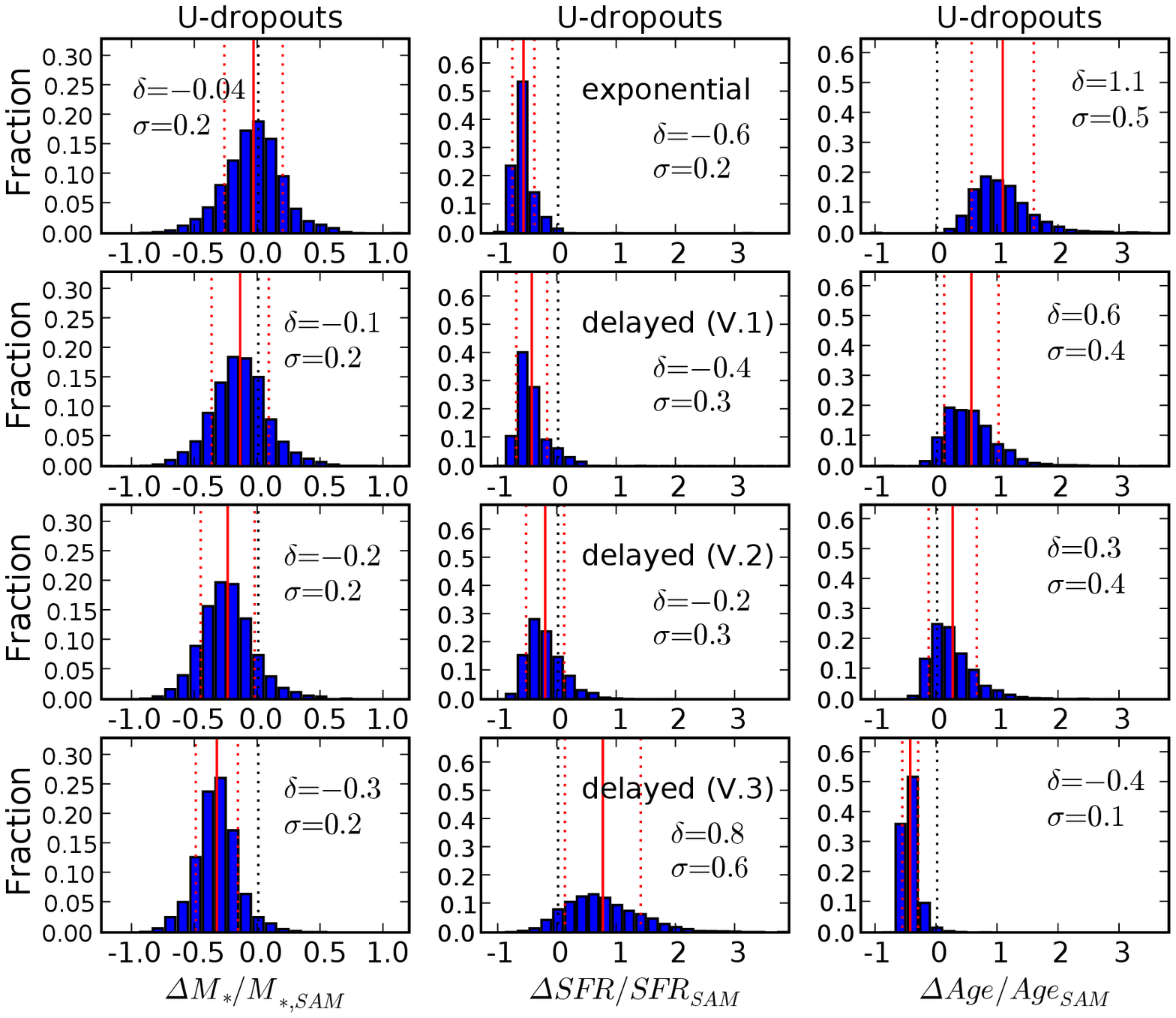}
\caption{Distributions of relative discrepancies in stellar-mass (left column), SFR  (middle column), and mean stellar population age (right column) estimation of $U$-dropouts.
Relative discrepancies are defined as \rm{$(value_{derived} - value_{intrinsic})/value_{intrinsic}$}.
The results are from the SED-fitting with delayed SFHs version 1 (the second row),
version 2 (the third row), and with linearly increasing SFHs (the fourth row).
The first row shows the results from the SED-fitting with exponentially declining SFHs from L09.
Red solid and dotted lines indicate the locations of the mean and the standard deviation of
$\Delta value / value_{intrinsic}$.
Black dotted lines show the location of $\Delta value / value_{intrinsic} = 0$.
Red solid and dotted lines show the locations of mean and standard deviation of relative discrepancies,
respectively.
Red dotted lines show the standard deviation of relative discrepancies.
In each panel, $\delta$ and $\sigma$ show the estimates of systematic offset and standard deviation,
respectively. \label{fig1}}
\end{figure}

\begin{figure}
\plotone{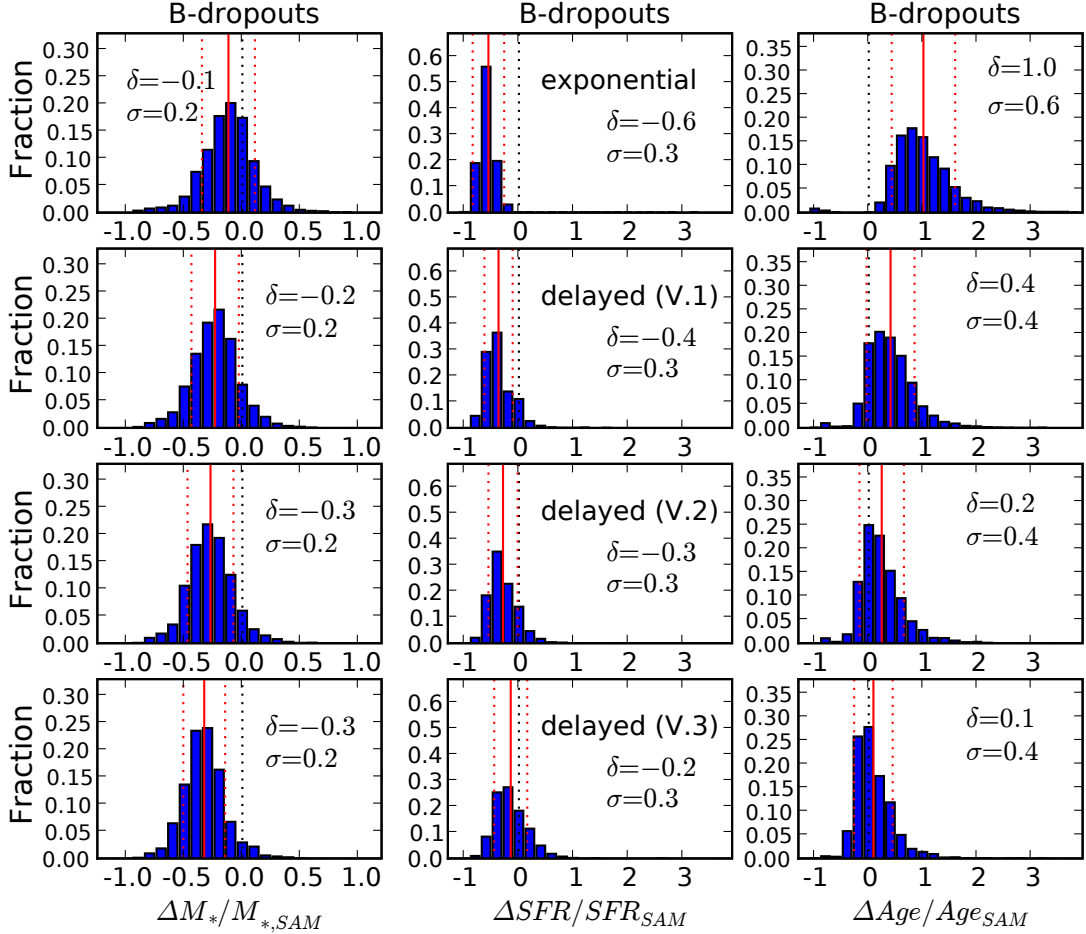}
\caption{Distributions of relative discrepancies in stellar-mass (left column), SFR (middle column), and mean stellar population age (right column) estimation of $B$-dropouts.
The results are from the SED-fitting with delayed SFHs version 1 (the second row),
version 2 (the third row), and linearly increasing SFHs (the fourth row).
The first row shows the results from the SED-fitting with exponentially declining SFHs from L09.
In each panel, $\delta$ and $\sigma$ show the estimates of systematic offset and standard deviation,
respectively.
Color- and type-coding of lines is the same as in Figure 1. \label{fig2}}
\end{figure}

\begin{figure}
\plotone{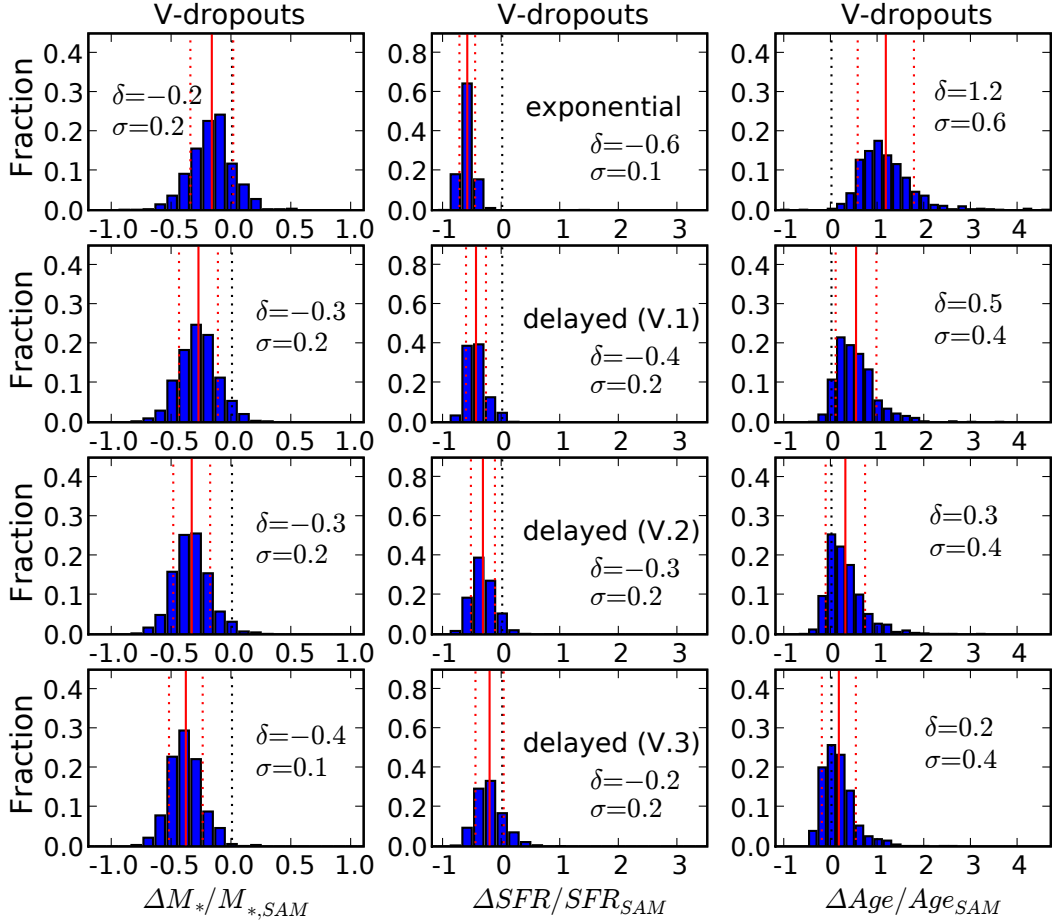}
\caption{Distributions of relative discrepancies in stellar-mass (left column), SFR (middle column), and mean stellar population age (right column) estimation of $V$-dropouts.
Details are similar with Figures 1 and 2. \label{fig3}}
\end{figure}

Figures \ref{fig1}-\ref{fig3} show the distributions of relative discrepancies
in the estimations of stellar masses, SFRs, and mean stellar population ages
(including the results from single-component fitting with exponentially
declining SFHs from L09).
In SED-fitting with exponentially declining SFHs, which is performed in L09,
$\tau$ (star formation time-scale parameter) is varied from 0.2 to 15.0 Gyr.
Relative discrepancies are defined as
$\rm{(value_{derived} - value_{intrinsic})/value_{intrinsic}}$, where
$\rm{value_{derived}}$ and $\rm{value_{intrinsic}}$ are SED-derived and intrinsic
stellar mass, SFR, or mean age.
In these figures, red solid (vertical) and black dotted lines show the
locations of mean relative discrepancies and of zero points, respectively ---
the distance between red solid and black dotted lines shows the amount of
systematic offset.
Red dotted lines show the standard deviation of relative discrepancies.

First, from these figures, we can see that stellar mass (left column) is the
least influenced by the assumed form of SFHs.
All three versions of delayed SFHs give similar stellar
mass distributions (but result in slightly more significant underestimation of
the stellar masses than exponentially declining SFHs).
When the exponentially declining SFHs are assumed in SED-fitting, the
relatively good recovery of stellar mass is largely fortuitous (see discussion
in L09's Section 4.3.2).
The mass-to-light ratio of the bulk of the population is clearly incorrect
because the ages are overestimated.
But because $\tau$-models are constrained to have lower instantaneous
SFRs than past SFRs, the SED-fitting tends
to make a compromise for the shortfall in young stars by boosting the total
number of stars.
In the delayed and rising SFR models, the ages and SFRs come out much closer
to the input values, but the stellar masses are now more sensitive to
the detailed mismatch between the simple parametric models and the true
SFHs because younger stars are dominating the SEDs.

Secondly, relative discrepancies for SFRs (middle column) and mean ages
(right column) derived from SED-fitting with delayed SFHs (the second, third, and
fourth rows) are on average smaller as well as less biased than the
distributions derived from SED-fitting with exponentially declining SFHs (the
first row; results from L09).
For $B$- and $V$-dropouts, the derived SFRs and mean ages show progressive
improvement from exponentially declining SFHs (the first row) to delayed SFHs
with 0.2 Gyr $< \tau <$ 2.0 Gyr (the second row) to ``$\tau \sim 0.9 \times t$"
delayed SFHs (the third row) to linearly increasing SFHs (the fourth row).
For $B$-dropouts, the mean relative discrepancy of SFR is reduced from $-56 ~\%$
to $-15 ~\%$, and it is reduced from $-60 ~\%$ to $-22 ~\%$ for $V$-dropouts.
The estimate of mean ages is more significantly improved and the mean relative
discrepancy is reduced from $100 ~\%$ to $9 ~\%$ for $B$-dropouts and from
$120 ~\%$ to $16 ~\%$ for $V$-dropouts.
In the case of $U$-dropouts, the ``$\tau \sim 0.9 \times t$" delayed SFHs are
shown to provide the best matches for SFR and age.
The mean relative discrepancy of SFR is reduced from $-58 ~\%$ to $-22 ~\%$
and the mean relative discrepancy of age is reduced from $110 ~\%$ to $26 ~\%$.
As can be speculated from Figures \ref{fig1}-\ref{fig3}, the amounts of biases
in SFR and age estimations are connected with each other --- in a sense that
improved match in stellar population ages leads to the reduced biases in SFR
estimation.

In Table \ref{tab1}, we list the means and standard deviations of relative
discrepancies in each physical parameter for SED-fitting with exponentially
declining SFHs and for SED-fitting with increasing SFHs ---
``$\tau \sim 0.9 \times t$" delayed SFHs for $U$-dropouts and linearly
increasing (i.e., ``$\tau$ = 10.0 Gyr") delayed SFHs for $B$- and $V$-dropouts.

The behavior of the derived dust reddening, $E(B-V)$, is similar
to that of the derived SFRs.
The reddening is significantly underestimated for the exponential
models and is closer to the true (spatially averaged, mean) reddening
of the semi-analytic models when the increasing SFR models are adopted.
However, the reddening values are still underestimated by about
$50 \%$ on average.

There is a difference in the treatment of metallicity between SED-fitting and
semi-analytic models.
In SED-fitting, any SED template from population synthesis models assumes a single
(discrete) value of metallicity, while mock LBGs from semi-analytic models are
composed of multiple stellar populations with different metallicity values ---
i.e., semi-analytic models accommodate continuous chemical enrichment history.
This difference would affect the SED-fitting results as well, but the best-fit
metallicity values derived from SED-fitting do not show any significant change
between SED-fitting with exponentially declining SFHs assumed and SED-fitting
with increasing (or delayed) SFHs assumed.
Therefore, metallicity does not significantly affect the changes in SED-derived
best-fit SFRs, ages, and stellar masses between SED-fitting with declining SFHs
and SED-fitting with increasing SFHs.
Detailed discussion about the effect of metallicity treatment (single, discrete
value versus continuous chemical enrichment) on the SED-fitting results is out
of the scope of this paper.

\subsection{Principal Component Analysis of the Star Formation Histories of Galaxies}

To understand better the sources of biases in SED-fitting, we analyze the
(mass-normalized) SFHs (i.e., SFHs divided by the final
stellar mass) of model $B$-dropouts using principal component analysis (PCA).

The PCA is, in principle, a statistical
method of linear transformation.
What PCA does is transforming a set of correlated variables into
a set of uncorrelated (i.e., orthogonal) variables --- which are
called principal components (PCs).
In the PCA, we generally want to reduce the
dimension (i.e., number of variables) in the data --- to simplify
the given problem or to find a pattern in the data --- without
losing critical information content of a given data set.
Therefore, the task is, in principle, to find a small number of
PCs which can account for as large fraction
of the variability of a given data set as possible.

This PCA has been widely used in many areas of
astronomy --- especially in characterizing and/or classifying
observed spectra of stars, galaxies, or quasars
\citep[e.g.,][]{bai98,van06,mad03}.
In this work, we use this widely used PCA method in an area, in which
this method has not been applied yet --- i.e., in analyzing and
classifying the SFHs of (mock) galaxies.
Through this PCA of SFHs, we reveal the
characteristic SFHs of model LBGs as well as classify the different types of SFHs.

We find that the first three PCs account for 99 $\%$ of the
variance in the SFHs for the $B$-dropouts.
From the PCA, we construct the mean (or representative) SFHs,
$\Psi_{rep}$ of total $B$-dropout galaxies by adding
the product of the PCs and the corresponding
mean weights (or eigenvalues) up to the fourth component:

\begin{equation}
\Psi_{rep} = \sum^{4}_{k=1} w_{k} \psi_{pc,k},
\end{equation} \label{eqn:repsfhs}

where $\psi_{pc,k}$ and $w_{k}$ are $k$th PC (i.e.,
eigen-SFH) and corresponding mean weight, respectively.

\begin{figure}
\plotone{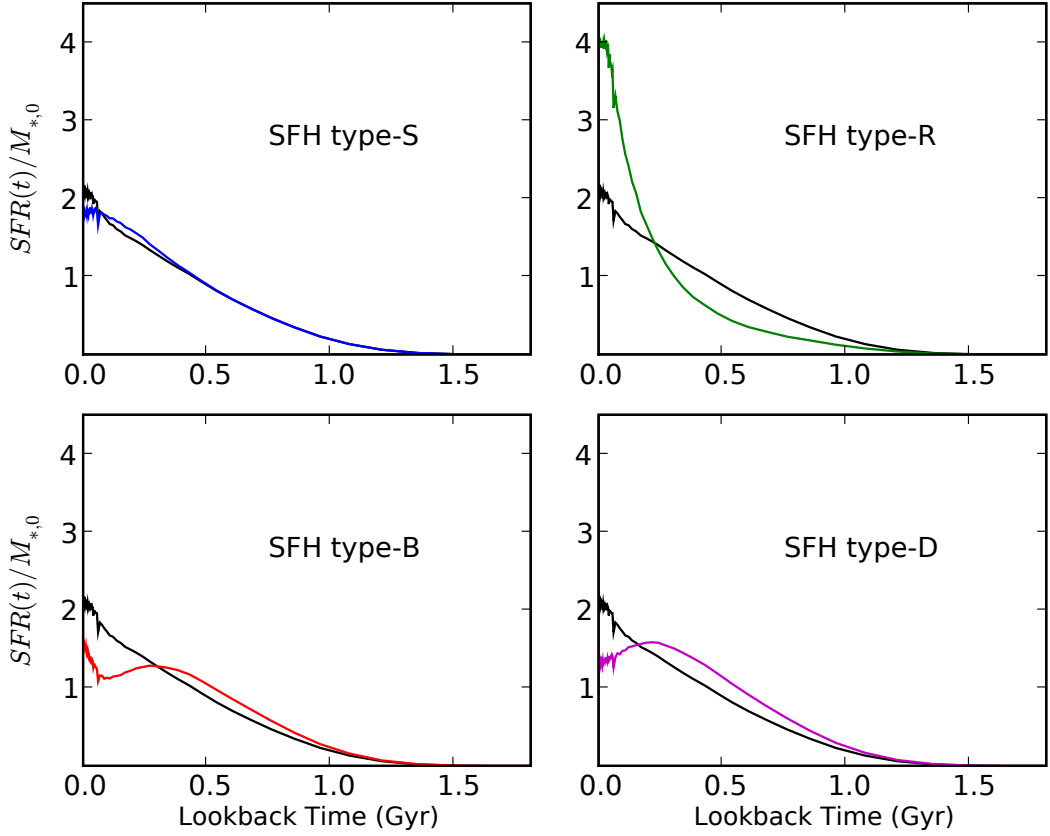}
\caption{Representative SFHs of SFH type-S (blue, upper left), of SFH type-R
(green, upper right), of type-B (red, lower left), and of type-D (purple, lower right) $B$-dropouts.
These representative SFHs are constructed by adding the products of PCs and
corresponding eigenvalues up to the fourth principal components derived from the PCA.
Please refer Section~3.2 for the definition of each SFH type.
The black line in each plot is the mean SFH of total $B$-dropouts. \label{fig4}}
\end{figure}

As shown via the black line in each panel of Figure \ref{fig4}, the mean SFH of
total $B$-dropouts is well represented by a linearly increasing
SFH \footnote{Even though not shown here, the mean SFHs of
$U$- and $V$-dropouts are also well represented by a linearly increasing SFH}.
This explains why the linearly increasing model improves the match in the age
and SFR distributions of $B$-dropouts.
By assuming linearly increasing SFHs, the difference between intrinsic SFHs
of SAM LBGs and the assumed forms of SFHs in SED-fitting has been reduced.
This results in reduced bias in age estimation (and SFR estimation).

\begin{figure}
\plotone{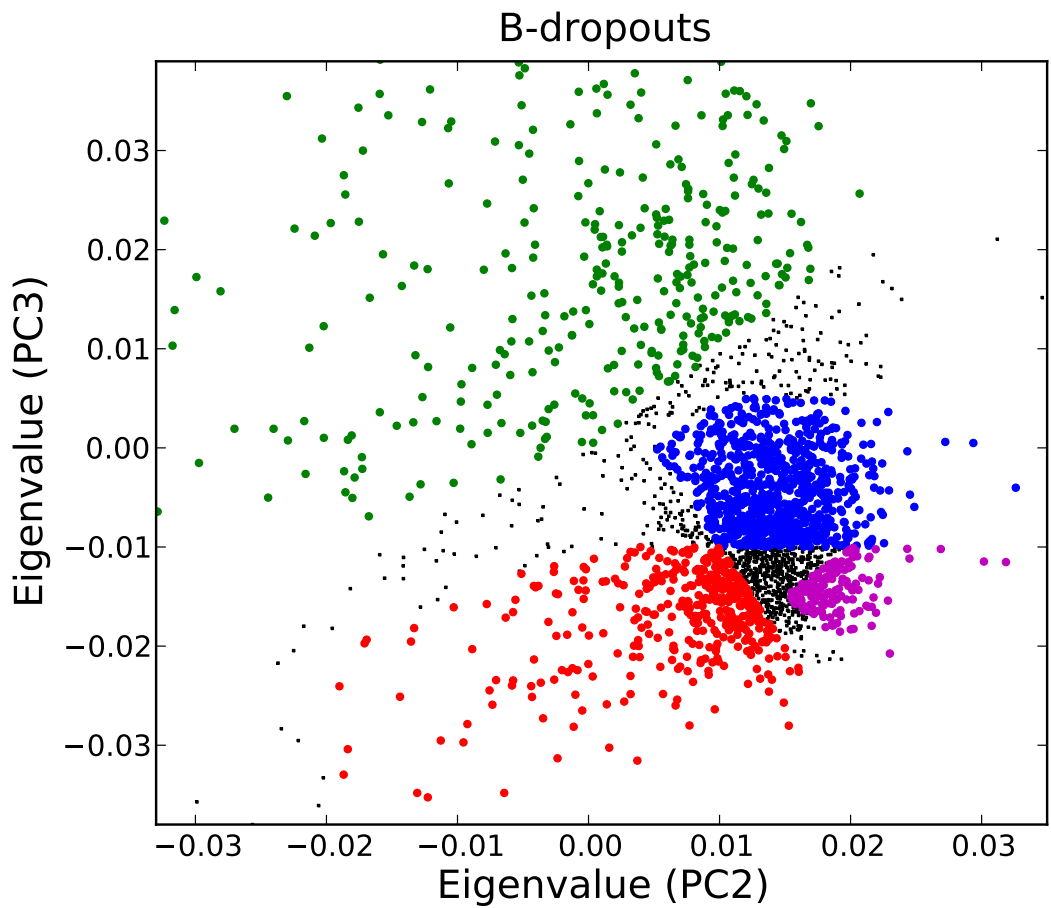}
\caption{Distribution of eigenvalues of the second ($x$-axis) PC
and of the third ($y$-axis) PC for $B$-dropouts LBGs.
Galaxies which belong to each SFH type is shown as
blue (type-S), green (type-R), red (type-B), and purple (type-D) points.
Black points represent the galaxies which do not belong to any SFH type --- having
either intermediate-type or peculiar SFHs. \label{fig5}}
\end{figure}

The different types of SFHs can be classified
through this PCA of SFHs.
The model LBGs' SFHs can be categorized based on the
eigenvalues (weights) of the second and the third PCs. The eigenvalues of
the first PC are similar for all types of galaxies and accounts for the
overall trend of SFHs of LBGs.
Distribution of the weight values corresponding to the second and the third
PCs is shown in Figure \ref{fig5}.
Here, blue points represent $B$-dropouts with type-S SFHs, green
points are for SFH type-R, red points are for type-B, and purple points
are the galaxies with type-D SFHs.
We name the four types of LBGs' SFHs as type-S
(Slowly increasing), type-R (Rapidly increasing), type-B
(having a Bump), and type-D (started to Decrease), following
the trend of each SFH type.
The colored lines in Figure \ref{fig4} show the mean
SFHs of four types of $B$-dropouts, constructed from the
PCs and eigenvalues.\footnote{The mean SFH
of each SFH type is constructed by adding the products of the PCs
and the corresponding mean eigenvalues of each SFH type up to the fourth component.}
For the $B$-dropout galaxies, about $30 \%$, $20 \%$, $20 \%$, and $6 \%$ of
LBGs are classified as type-S, type-R, type-B, and type-D, respectively.
The remaining $B$-dropouts show intermediate types of SFHs.

From Figure \ref{fig5}, we can see that the weight --- i.e., eigenvalue --- of
the third PC distinguishes the SFH type-R from other SFH types, and
that the weight of the second PC makes the difference between the type-B
and the type-D LBGs.
Not all LBGs can be clearly classified into one of these four SFH types.
Some galaxies have peculiar SFHs and some LBGs show
SFHs which are intermediate between different SFH types.
For example, black points in the region between type-S, type-B, and
type-D represent the galaxies with intermediate SFHs.
%
%
%

\begin{figure}
\plotone{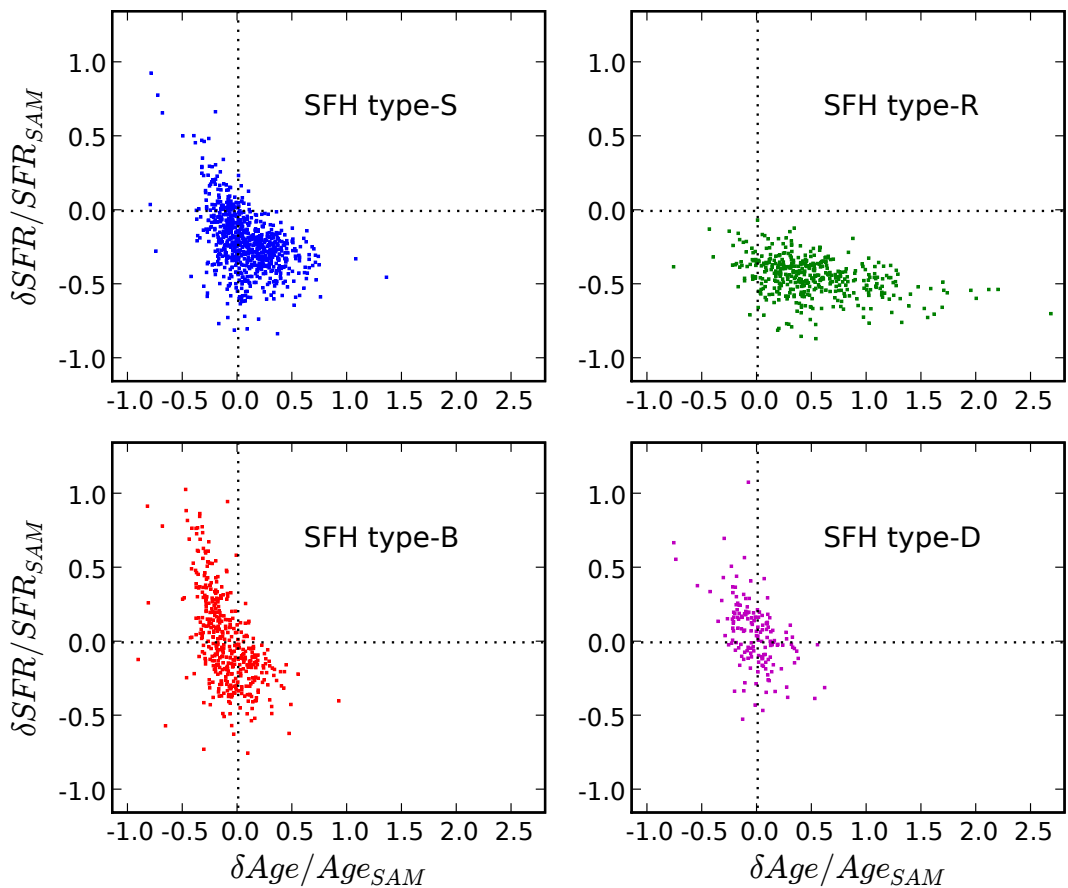}
\caption{Relative SFR discrepancy vs. relative age discrepancy of SFH type-S (blue, upper left),
type-R (green, upper right), type-B (red, lower left), and type-D (purple, lower right) model $B$-dropouts
in SED-fitting with increasing SFHs.
Please refer Section 3.2 for the definition of each SFH type. \label{fig6}}
\end{figure}

Next, we analyze the behaviors of biases for each different SFH type, revealing
that the pattern and amount of bias in SED-fitting are strongly dependent
on the intrinsic SFHs of LBGs.
Aspects of these SFH-dependent biases are shown in Figure \ref{fig6}, where we
plot $B$-dropout galaxies with each SFH type in the ``relative age error - relative
SFR error" space in case of SED-fitting with linearly increasing SFHs --- i.e.,
delayed SFHs with $\tau = 10.0$ Gyr.
This figure shows that stellar population ages are mostly
overestimated for SFH type-R galaxies despite the relatively good
match for each entire dropout sample.
SFRs are mostly underestimated for these R-type
galaxies\footnote{Although not
shown here, the two-component fitting --- with a young burst
embedded in an older, slowly varying component ---  which is performed in L09 ---
shows smallest bias in SFR estimation for type-R galaxies (but not for other
types of galaxies).}.
If we compare SFH type-S and type-D galaxies, we can see that
biases in SFR estimation are larger (i.e., SFRs are
more likely underestimated) for type-S galaxies than type-D
galaxies while the amount of biases in age estimation is similar for
these two types of galaxies.

\section{Estimation of Stellar Mass and Star Formation Rate from Rest-frame Optical and UV Luminosity} \label{calib}

An obvious question is whether SED fitting performs better than simple heuristic
estimates using some estimate of color and luminosity.
To explore this question, in this section, we derive (1) stellar masses of LBGs
from rest-frame optical luminosity and UV-optical color and
(2) SFRs from rest-frame UV color and UV luminosity.
We find that these simple estimates perform almost as well as SED fitting with a
rising SFH (and that both techniques perform better than SED fitting with an
exponentially decaying SFH in the estimates of SFR and age).

First, we derive the calibration of stellar mass from rest-frame optical
luminosity (from IRAC 3.6 $\mu$m) and rest-frame UV--optical color (from
$J - m_{3.6 \mu m}$), using model galaxies from semi-analytic
model runs, as

\begin{equation} \label{op2ms}
M_{*} ~(M_{\odot}) = 8.03 \times 10^{-21} \times L_{\nu,7200} \times 10^{f(J-m_{3.6 \mu m},z)}.
\end{equation}

Here, $L_{\nu,7200}$ is (bolometric-correction included) specific luminosity at
rest-frame 7200 $\rm{\AA}$ (in unit of $erg ~s^{-1} ~Hz^{-1}$), and is
derived from IRAC 3.6 $\micron$.
$f(J-m_{3.6 \mu m},z)$ is a function of ($J-m_{3.6 \mu m}$) color and of redshift, $z$,
and is given by

\begin{equation} \label{fcolor}
f(J-m_{3.6 \micron},z) = 0.606 \times (J-m_{3.6 \mu m}) + 0.297 - 0.0749 z,
\end{equation}
within the redshift range of ($2.7 \leq z \leq 5.6$).

Here, the color dependence of $f(J-m_{3.6 \micron},z)$ comes from the variation of
galaxies' mass-to-light ratio with parameters such as age, extinction, and metallicity.
The redshift-dependent term is the $k$-correction and is derived from BC03 spectral
templates with linearly increasing SFHs.

Next, we estimate SFRs from rest-frame UV luminosity and rest-frame
UV color, again using model galaxies from semi-analytic model runs, as

\begin{equation} \label{uv2sfr}
\Psi ~(M_{\odot} \rm{yr^{-1}}) = 5.22 \times 10^{-29} \times L_{\nu,1600} \times 10^{0.4 E(B-V) k(\lambda)}.
\end{equation}

Here, $L_{\nu,1600}$ is specific luminosity at rest-frame 1600 $\rm{\AA}$ (in unit of
$\rm{erg ~s^{-1} ~ Hz^{-1}}$), and is derived from ACS $i_{775}$ for $U$- and $B$-dropouts
and from ACS $z_{850}$ for $V$-dropouts with the appropriate bolometric correction.
The color excess, $E(\bv)$, is calibrated as a function of rest-frame UV color, which is a
photometric measure of spectral slope, and redshift, and is given as

\begin{eqnarray} \label{ebmv}
E(B-V)&=& 0.448 \times (V_{606} - i_{775}) + 0.618 - 0.178 z ~(2.8 \leq z \leq 3.8)\\
&=& 1.28 \times (i_{775} - z_{850}) - 0.108 + 0.0678 z ~(3.5 \leq z \leq 4.6)\\
&=& 0.789 \times (z_{850} - J) + 0.200 - 0.00762 z ~(4.4 \leq z \leq 5.6),
\end{eqnarray}
and $k(\lambda)$ is the reddening curve from \citet{cal00} (Equation (4)).
The redshift dependence of $E(\bv)$ accounts for the $k$-correction and is derived from
BC03 spectral templates with linearly increasing SFHs.

Under the assumption of linearly increasing SFHs, we can estimate the mass-weighted
mean age from the estimated values of SFR and stellar mass:

\begin{equation} \label{agecalc}
\rm{Age ~(Gyr)} = \frac{2}{3} \times T(M_{*},\Psi),
\end{equation}
where $T(M_{*},\Psi)$ is $(M_{*} (M_{\odot}))/(\Psi (M_{\odot} yr^{-1}) \times 10^{9})$.

With linearly increasing SFHs, $T(M_{*},\Psi)$ must be within the range of
$0.2 \leq T(M_{*},\Psi) \leq 0.9$ because the value of $t$ in Equation(1) ---
time since the onset of star formation --- should be smaller than the age
of the universe at corresponding redshift.
Values of $T(M_{*},\Psi)$ outside of this range can result from a deviation of the
actual SFHs from linearly increasing SFHs or from the propagation
of errors in stellar mass or SFR estimate.
If $T(M_{*},\Psi) < 0.2$, we estimate the mean age as $T(M_{*},\Psi)$ with the assumption
of $\Psi \varpropto t^{\alpha}$ with $\alpha \gg 1$.
If $0.9 < T(M_{*},\Psi) < 1.5$, the mean age is calculated as $0.5 \times T(M_{*},\Psi)$
assuming constant SFH, and if $T(M_{*},\Psi) > 1.5$, the mean age is $\frac{1}{3} \times T(M_{*},\Psi)$
assuming $\Psi \varpropto t^{-1/2}$.

The distributions of relative discrepancies in estimation of stellar mass, SFR, and mean
age using the calibrations explained in this section are shown in Figure \ref{fig7} for
$U$-dropouts (top row),  $B$-dropouts (middle row), and $V$-dropouts (bottom row).
As can be seen in Figure 7, the calibrations introduced here also provide better
estimation of SFRs and mean ages than SED-fitting with exponentially declining SFHs.
Mean relative discrepancies of SFR are --0.32, 0.29, and 0.05 for $U$-, $B$-, and $V$-dropouts,
respectively, and mean relative discrepancies of age are 0.10, --0.20, and --0.18.
Here, we use exact redshift information in derivation of these properties.
If we use the mean redshift for each $U$-, $B$-, and $V$-dropout sample, the discrepancy
between the derived distribution and the intrinsic one increases slightly for SFR and
age, while there is no noticeable change in the stellar mass distribution.

\section{Conclusion}

In this paper, we extend our previous effort to examine how well the widely used
SED-fitting methods can recover the intrinsic distributions of physical parameters
--- stellar mass, SFR, and mean age --- of high-redshift, star-forming galaxies.
Here, we perform SED-fitting analysis of model LBGs ($3.0 \lesssim z \lesssim 5.6$)
with ``delayed" SFHs assumed.
The main results of this paper are as follows: (1) for SFR and mean-age distributions,
increasing SFHs provide a better match than exponentially decreasing SFHs
or other types of delayed SFHs, (2) stellar masses are slightly more underestimated
with linearly increasing SFHs, and (3) the behaviors of biases strongly depend on
galaxies' intrinsic SFH types.
While the stellar mass estimates from the $\tau$-models appeared to be
more ``robust", this was largely a fortuitous cancellation of the errors in
SFR and the errors in the mean mass-to-light ratios of the older stars.
The rising SFH models do worse than the $\tau$-models in estimating stellar mass but we
should regard this as a challenge for future ``simple models" rather than a major failing.
Because these new models have stars that are on average much younger --- and much closer
to the correct ages of the input models --- the stellar masses are much more sensitive
to the actual age distribution.
In SED-fitting with $\tau$-models, stellar masses do not change dramatically with
an adopted value of $\tau$ or with an adopted time for the initial burst provided
it was a long time ago.
These new models are much more sensitive to the detailed shape of the star-forming
history, which is still not quite right (on average).

In this paper, we present the results of the SED-fitting with delayed/increasing SFHs
for LBGs from semi-analytic models.
We have also analyzed the SEDs of observed LBGs ($3 \leq z \leq 5$) found in
the GOODS-S field, and the results show that the increasing SFHs provide better
results (than exponentially decaying SFHs) for the observed LBGs, too.
We will present the SED-fitting analysis results for GOODS-S LBGs separately.

We analyze the SFHs of SAM LBGs using PCA, showing that the SFRs
of LBGs in modern galaxy evolution models tend to increase with time over
the redshift range $3 < z < 6$.
This provides the reason why assuming increasing SFHs provides better estimates
of stellar population ages (and SFRs) of LBGs in SED-fitting than when
widely-used exponentially declining ones are assumed.
Using PCA, we can also classify the SFHs of LBGs, and show SFH-dependent behavior
of biases in SED-fitting.

\begin{figure}
\plotone{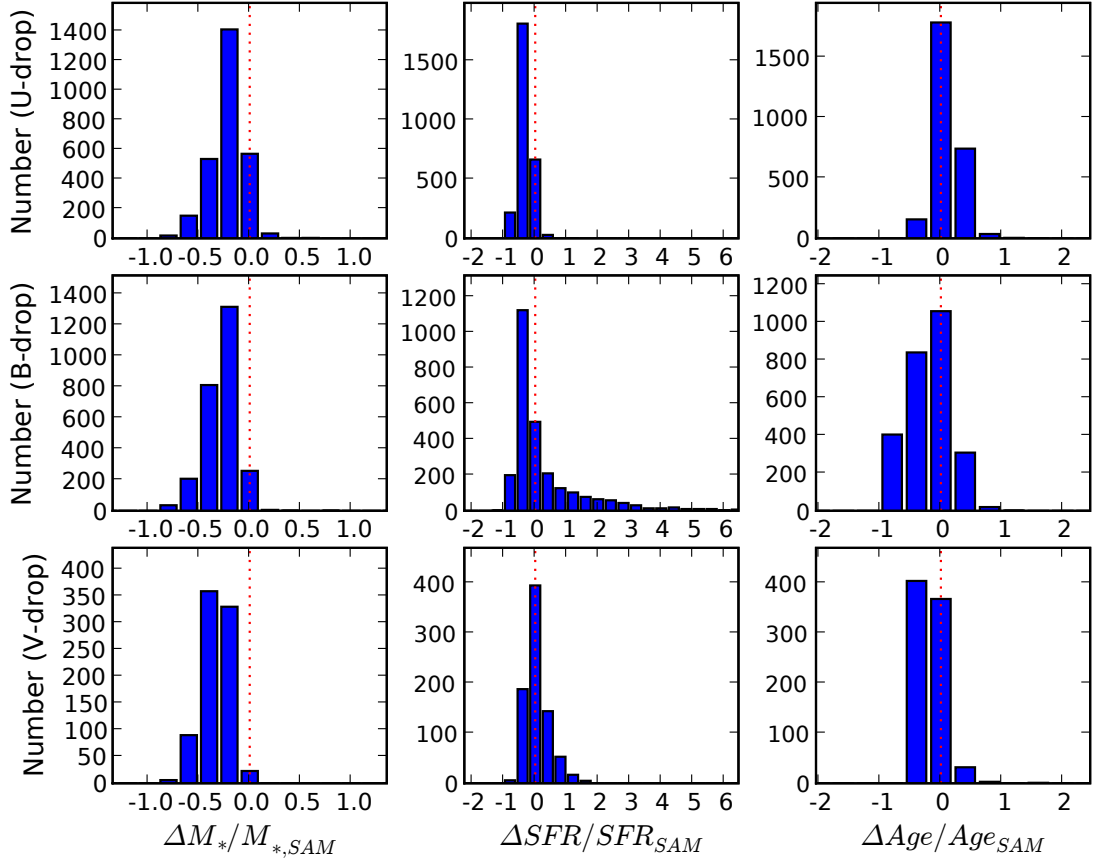}
\caption{Distributions of relative discrepancies in stellar-mass (left column), SFR (middle column), and mean stellar population age (right column) estimation of $U$-dropouts
(the first row), $B$-dropouts (the second row), and $V$-dropouts (the third row).
Estimation of these physical parameters is done using the calibrations explained in
Section 4. \label{fig7}}
\end{figure}

We also present the calibrations from broadband photometry to stellar mass, SFR,
and mean age in Section~\ref{calib}.
As shown in Figure \ref{fig7}, these calibrations provide estimates
for stellar mass, SFR, and mean age that are nearly as good as the estimates from
SED-fitting with linearly increasing SFHs.


Biases arising in SED-fitting can have several effects on the investigation
of galaxy evolution.
As shown in this paper, the biases in SFR estimation depend on the intrinsic
SFHs of galaxies.
Because the intrinsic stellar masses of LBGs also depend on their SFHs
(e.g., type-D galaxies are, on average, more massive than type-R galaxies),
this can cause biases in measurement of the slope (also the normalization or zero point,
especially in case of SED-fitting with exponentially declining SFHs) of
``stellar-mass - SFR" relation or ``stellar-mass - SSFR (SFR per unit stellar mass)"
relation.

The results of this paper --- e.g., improvement in SFR- and age-estimation with linearly
increasing SFHs --- hold for star-forming galaxies at $3 \lesssim z \lesssim 6$.
Interesting question is how different results we will get at lower redshift ($z < 3$)
or for different galaxy populations --- e.g., old galaxies with little on-going
star formation or red infrared-luminous galaxies.

\acknowledgments

This research was jointly supported through the GOODS and COSMOS survey programs,
NASA grants HST-GO-09822 and HST-GO-09425, the Spitzer Space Telescope Legacy Science
Program, provided by NASA contract 1224666 issued by the JPL, Caltech, under NASA
contract 1407.

\clearpage

\clearpage

\clearpage

\clearpage

\clearpage

\clearpage

\clearpage

\clearpage

\begin{deluxetable}{ccccc}
\tablecolumns{5} \tablewidth{0pc} \tablecaption{Systematic Offset
and Random Scatter of Physical Parameters in SED-fitting
\label{tab1}}
\tablehead{ \colhead{Parameter}   &
\colhead{$\delta$\tablenotemark{a} (Decaying SFH)}   &
\colhead{$\sigma$\tablenotemark{b} (Decaying SFH)}   &
\colhead{$\delta$ (Increasing SFH\tablenotemark{c})}   &
\colhead{$\sigma$ (Increasing SFH)}} \startdata
\sidehead{$U$-dropouts}
Stellar mass & -0.04 & 0.23 & -0.24 & 0.21 \\
SFR & -0.58 & 0.19 & -0.22 & 0.32 \\
Age & 1.1 & 0.51 & 0.26 & 0.39 \\
\sidehead{$B$-dropouts}
Stellar mass & -0.12 & 0.23 & -0.33 & 0.18 \\
SFR & -0.56 & 0.29 & -0.15 & 0.30 \\
Age & 1.0 & 0.59 & 0.09 & 0.36 \\
\sidehead{$V$-dropouts}
Stellar mass & -0.17 & 0.18 & -0.39 & 0.14 \\
SFR & -0.60 & 0.13 & -0.22 & 0.24 \\
Age & 1.2 & 0.60 & 0.16 & 0.36 \\
\enddata

\tablenotetext{a}{$\delta$ is the mean value of relative discrepancies and is a measure of systematic offset.}
\tablenotetext{b}{$\sigma$ is the standard deviation of relative discrepancies.}
\tablenotetext{c}{Here, increasing SFHs are ``$\tau \sim 0.9 \times t$" delayed SFHs for $U$-dropouts, and
linearly increasing (i.e., ``$\tau$ = 10.0 Gyr" delayed) SFHs for $B$- and $V$-dropouts.}

\end{deluxetable}


\begin{thebibliography}{}

\bibitem[Bailer-Jones et al.(1998)]{bai98} Bailer-Jones, C. A. L., Irwin, M., \& von Hippel, T. 1998, \mnras, 298, 361
\bibitem[Bruzual \& Charlot(2003)]{bru03} Bruzual, G. \& Charlot, S. 2003, \mnras, 344, 1000
\bibitem[Calzetti et al.(2000)]{cal00} Calzetti, D., Armus, L., Bohlin, R. C.,
Kinney, A. L., Koornneef, J., \& Storchi-Bergmann, T. 2000, \apj, 533, 682
\bibitem[Chabrier(2003)]{cha03} Chabrier, G. 2003, \pasp, 115, 763
\bibitem[Conroy et al.(2009)]{con09} Conroy, C., Gunn, J. E., \& White, M. 2009, \apj, 699, 486
\bibitem[Idzi et al.(2004)]{idz04} Idzi, R., Somerville, R., Papovich, C., Ferguson, H. C.,
Giavalisco, M., Kretchmer, C., \& Lotz, J. 2004, \apj, 600, L115
\bibitem[Lee et al.(2009)]{lee09} Lee, S. -K., Idzi, R., Ferguson, H. C., Somerville, R. S.,
Wiklind, T., \& Giavalisco, M. 2009, \apjs, 184, 100 (L09)
\bibitem[Madau(1995)]{mad95} Madau, P. 1995, \apj, 441, 18
\bibitem[Madgwick et al.(2003)]{mad03} Madgwick, D. S. et al. 2003, \apj, 599, 997
\bibitem[Maraston et al.(2006)]{mar06} Maraston, C. et al. 2006, \apj, 652, 85
\bibitem[Papovich et al.(2001)]{pap01} Papovich, C., Dickinson, M., \& Ferguson, H. C.
2001, \apj, 559, 620
\bibitem[Sawicki \& Yee(1998)]{saw98} Sawicki, M. \& Yee, H. K. C. 1998, \aj, 115, 1329
\bibitem[Shapley et al.(2001)]{sha01} Shapley, A. E., Steidel, C. C., Adelberger, K. L.,
Dickinson, M., Giavalisco, M., \& Pettini, M. 2001, \apj, 562, 95
\bibitem[Somerville \& Kolatt(1999)]{som99} Somerville, R. S. \& Kolatt, T. S. 1999, \mnras,
305, 1
\bibitem[Somerville et al.(2001)]{som01} Somerville, R. S., Primack, J. R., \& Faber, S. M.
2001, \mnras, 320, 504
\bibitem[Vanden Berk et al.(2006)]{van06} Vanden Berk, D. E. et al. 2006, \aj, 131, 84

\end{thebibliography}
\end{document}